\documentclass[12pt,british]{article}
\usepackage{mfirstuc}
\usepackage[T1]{fontenc}
\usepackage[latin9]{inputenc}
\usepackage{geometry}
\usepackage{textcomp}
\usepackage{graphicx}
\usepackage{setspace}
\usepackage{natbib}
\setcitestyle{authoryear,open={(},close={)}} 
\usepackage{booktabs}
\usepackage{caption}
\usepackage{hanging}
\usepackage{lscape}
\usepackage{afterpage}
\usepackage{enumitem}
\usepackage{float}
\makeatletter

\providecommand{\tabularnewline}{\\}

\usepackage{hyperref}
\usepackage{amssymb,amsfonts,amsmath}
\hypersetup{colorlinks=true,citecolor=blue,linkcolor=blue,urlcolor=blue}
\date{\today}

\makeatother

\usepackage{babel}
\title{Job Satisfaction Through the Lens of Social Media: 

Rural--Urban Patterns in the U.S.}

\author{Stefano M. Iacus\thanks{Harvard University, Institute for Quantitative Social Science, Cambridge (MA), USA. E-mail: \texttt{siacus@iq.harvard.edu}.} \and 
Giuseppe Porro\thanks{Department of Law, Economics and Culture, University of Insubria, Como, Italy. E-mail: \texttt{giuseppe.porro@uninsubria.it}.}}

\begin{document}

\maketitle

\begin{abstract}
We analyze a novel large-scale social-media-based measure of U.S. job satisfaction, constructed by applying a fine-tuned large language model to 2.6 billion georeferenced tweets, and link it to county-level labor market conditions (2013-2023). Logistic regressions show that rural counties consistently report lower job satisfaction sentiment than urban ones, but this gap decreases under tight labor markets. In contrast to widening rural-urban income disparities, perceived job quality converges when unemployment is low, suggesting that labor market slack, not  income alone, drives spatial inequality in subjective work-related well-being.
\end{abstract}



\section{Introduction}
The growing literature on subjective well-being (SWB) has increasingly examined the U.S.\ rural--urban divide, creating a research stream that spans psychology, sociology, economics, and political science.
In addition to studies on differences in life satisfaction \citep{atherton2024}, scholars have examined the relationship between local context and political orientation, which has in turn triggered a lively public debate (see, e.g., \cite{gimpel2024, brown2024}).

Less effort has been devoted to economic aspects of SWB, particularly those related to the labor market, i.e. job satisfaction, likely due to the lack of availability of fine-grained nationwide data. In fact, research on disparities in job satisfaction between U.S. rural and urban labor markets has relied, at best, on state-level small-sample surveys (\cite{martinson1984}). Other studies have examined satisfaction within specific worker categories (\cite{flores2020}; \cite{vick2016}; \cite{fossum1974}) or investigated particular aspects of employment, such as remote work (\cite{paul2022}). More recently, scholarly attention has shifted toward objective measures of job quality (\cite{carnevale2024}), rather than subjective satisfaction.

We use a new social media based indicator to study job-related satisfaction among rural and urban counties in the United States between 2013 and 2023, and relate it to well-documented trends in U.S. labor markets. Despite some apparently paradoxical patterns in subjective indicators, our results are broadly consistent with Autor's interpretation of recent urban versus rural labor-market dynamics \citep{autor2019}.

\section{U.S. Job satisfaction 2013--2023}
As noted above, the limited prior studies did not reach conclusive results regarding differences in job satisfaction across U.S. rural and urban contexts. At most, some evidence suggests a possible non-linear relationship between the degree of urbanization and job satisfaction, with individuals born in small towns reporting higher levels of satisfaction than those from open country, villages, or larger cities (\cite{martinson1984}).

Our new dataset enables a nationwide county-level assessment of SWB dimensions -- including job satisfaction -- since 2013 to 2023, a decade characterized by a progressive decline in unemployment rates, with the exception of the pandemic period (\cite{BLS2025}).

\subsection{The dataset}\label{sec:data}
The job  satisfaction social media indicator is part of a larger set of indicators from the project \textit{The Human Flourishing Geographic Index} (HFGI) \citep{iacus2025},  
conceptually inspired by the Harvard's Human Flourishing Program \citep{VanderWeele2017}, which defines flourishing  across six areas of well-being: happiness and life satisfaction, mental and physical health, meaning and purpose, character and virtue, close social relationships, and material and financial stability. The HFGI dataset offers high-resolution indicators at county and state level for the United States at both monthly and yearly frequency. It has been obtained  analyzing approximately 2.6 billion geo-referenced tweets from the Harvard CGA Geotweet Archive \citep{Lewis16}, classified using a  fine-tuned large language model (LLM) \citep{finetuning2024}. The dataset spans January 2013 to June 2023. 
The details on how these indicators have been computed through generative AI analysis are given in the Supplementary Material as well as in \cite{finetuning2024, iacus2025}.

The social media indicator \texttt{jobsat}  has been aggregated at county level. Its values vary in the interval $[-1,1]$. The indicator comes with the counts of tweets used to evaluate it, i.e., those tweets containing expressions of satisfaction or dissatisfaction with the job situation. This variable is called \texttt{TweetVol} in models~\eqref{eq:logit} and \eqref{eq:ols} below.

In addition to our social media indicator, we provide a set of control variables. 

We make use of the 2023 Rural-Urban Continuum Codes (\texttt{rural})  that distinguish U.S. metropolitan (metro, values 1-3) counties by the population size of their metro area, and non-metropolitan (nonmetro, values 4-9) counties by their degree of urbanization and adjacency to a metro area \citep{USDA_ERS_2024}. 

To control for county-level economic conditions, we include a measure of household income derived from the American Community Survey (ACS) 5-Year Estimates (\texttt{acs5}), which provides reliable small-area averages based on multi-year data \citep{uscb_acs5}.

Monthly county-level labor-market indicators are obtained from the U.S.\ Bureau
of Labor Statistics (BLS) \citep{BLSLAUS} Local Area Unemployment Statistics (LAUS) program via
the public BLS API \citep{BLSAPI}. For each county, we collected  non-seasonally-adjusted monthly unemployment rates and labor force data for period
2013-2023, as indicators of local labor-market conditions (named \texttt{UnemploymentRate} and \texttt{LaborForce} respectively in models~\eqref{eq:logit} and \eqref{eq:ols} below). To control for seasonality, we include monthly and yearly fixed effects in the models. 

\begin{figure}[ht]
 \caption{Monthly national unemployment rate (\%) and labor force (in million) obtained through county-level monthly aggregation compared to the social media indicator \texttt{jobsat} rescaled to national level to account for the missing counties in some years/months.}
    \label{fig:unemp_lforce}
    \centering
    \includegraphics[width=0.75\linewidth]{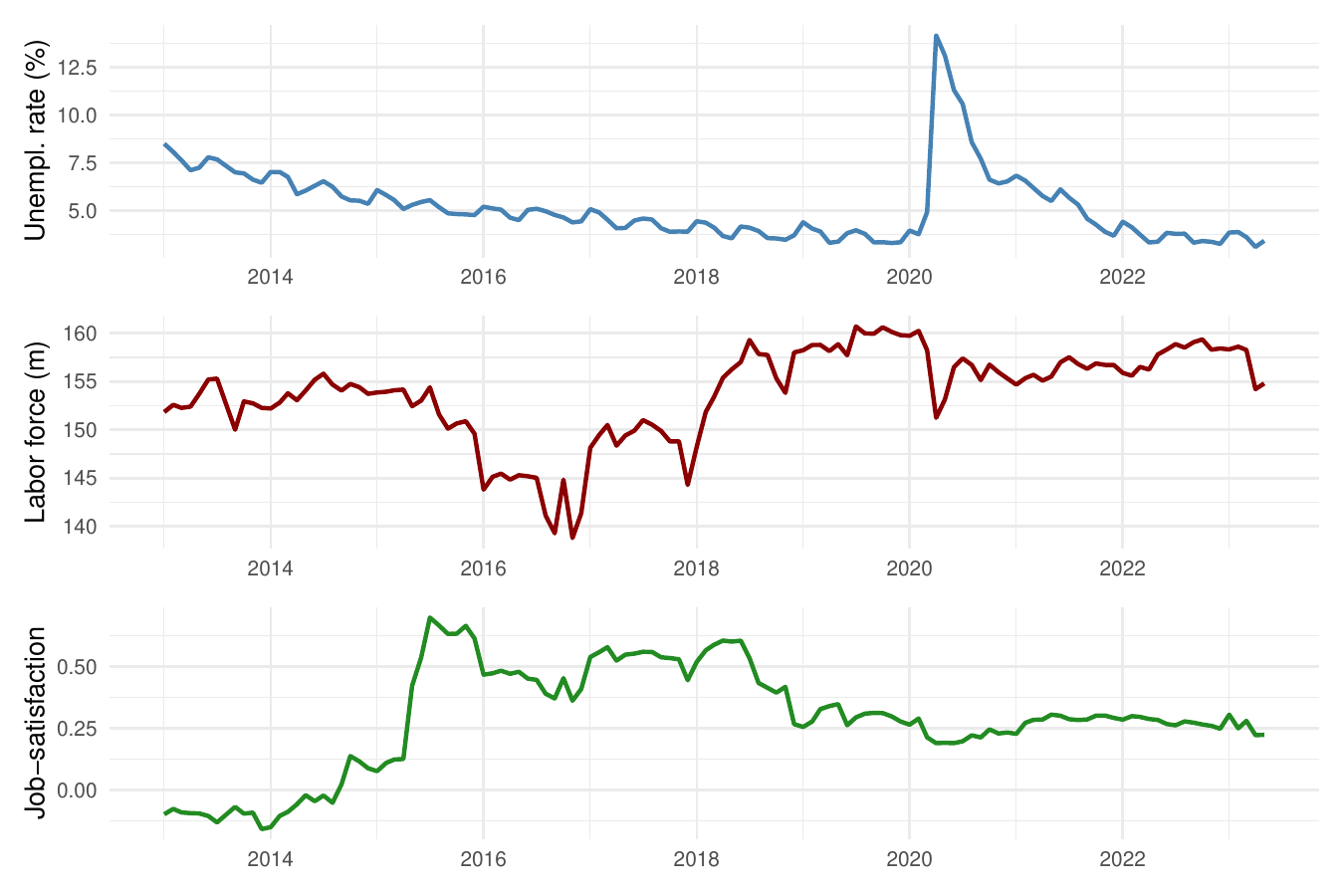}
\end{figure}

Figure~\ref{fig:unemp_lforce} plots the monthly national unemployment rate and labor force, which are constructed by aggregating county-level data, together with a version of the \texttt{jobsat} indicator that has been rescaled to reflect monthly variation in the number of contributing counties. This adjustment is applied only for visualization. The econometric analyses in Section~\ref{sec:model}  rely on the unmodified county-level indicators.

\begin{figure}[!ht]
    \centering
    \includegraphics[width=0.8\textwidth]{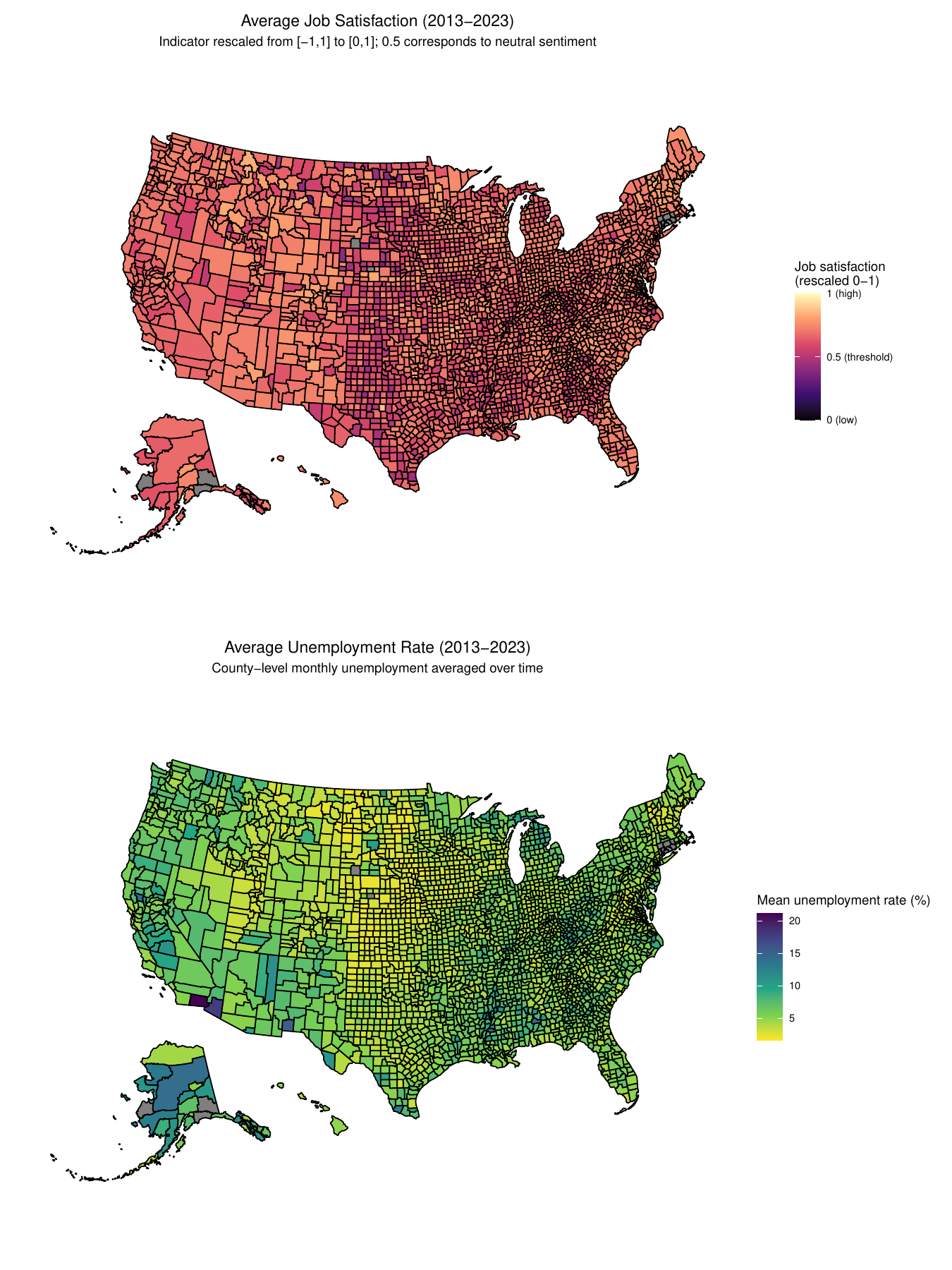}
    \caption{
        County-level averages of job-satisfaction sentiment (rescaled to the interval [0,1]) and unemployment rates in the United States, 2013--2023. 
        The top panel reports the rescaled \texttt{jobsat} indicator based on social-media sentiment, mapped using a perceptually uniform viridis color scale; higher values indicate more positive job-related sentiment.
        The bottom panel shows the mean county-level unemployment rate over the same period, using an inverted viridis  color scale so that brighter areas correspond to lower unemployment. 
        Counties with missing data in specific years contribute only when observed; maps reflect the average over available months for each county.
    }
    \label{fig:jobsat_unemployment_maps}
\end{figure}

Figure~\ref{fig:jobsat_unemployment_maps} compares the spatial distribution of 
(i) the average \texttt{jobsat} indicator rescaled from its original $[-1,1]$ range to $[0,1]$ to facilitate the visual interpretation, and 
(ii) the average unemployment rate, both computed at the county level over the 2013--2023 period. 
For each county, we first aggregate monthly observations into a single time-averaged value, noting that the set of counties contributing data varies slightly across years depending on tweet availability and BLS reporting. 
The maps therefore depict long-run spatial patterns based on all available observations for each county. 
The use of perceptually uniform viridis color scales ensures comparability across panels. 
Together, the two maps provide a spatial overview of baseline labor-market conditions and expressed job-related sentiment, which serve as the foundation for the econometric analysis presented  below.

\subsection{A Rural--Urban Disparity?}\label{sec:model}

To investigate the determinants of job-satisfaction sentiment (our \texttt{jobsat} indicator), we estimate a sequence of four nested logistic regression models that progressively incorporate additional socioeconomic and contextual covariates. We introduce a binary county-month outcome variable
$Y_{itm}=\textbf{1}_{\{\texttt{jobsat}_{itm}>0\}}$, which equals one if the average job-satisfaction score in county 
$i$ at time $(t,m)$ is positive and zero otherwise. Model~(1) includes the county--month unemployment rate (\texttt{UnemploymentRate}) as the main explanatory variable, together with the size of the county labor force (in thousands, \texttt{LaborForce}), number of tweets used to calculate the indicator (\texttt{TweetVol}), and year and month fixed effects. We use labor force rather than population to better capture the economic relevance of counties of different sizes: the labor force reflects the share of residents directly exposed to labor market conditions, which is more theoretically aligned with job-related sentiment than total population.
Model~(2) augments this baseline specification with the ACS 5-year income estimate (\texttt{ACS5}), capturing longer-term economic well-being. Model~(3) introduces an interaction between the unemployment rate and labor force. This interaction allows the marginal association between unemployment and job-satisfaction sentiment to vary with the economic scale of the county: in small labor markets, changes in unemployment may carry different informational or economic weight than in large ones.
Finally, Model (4) adds rurality (\texttt{Rural}) to capture spatial variation in socioeconomic context, and introduces state fixed effects to absorb persistent cross-state heterogeneity. All models include year ($\gamma$) and month ($\delta$) fixed effects to absorb  temporal shocks and seasonal patterns. The fully specified Model~(4), which forms the basis of our interpretation and predictive analyses, is given in equation~\eqref{eq:logit} and incorporates  state fixed effects ($\theta_{s}$) to account for national specific characteristics. Estimated coefficients for all four models are reported in Table~\ref{tab:jobsat}, with fixed effects omitted for brevity. These fixed effects follow standard panel-data practice and help control for unobserved temporal heterogeneity arising from macroeconomic shocks (e.g., the COVID-19 pandemic), national policy changes, and other broad influences affecting all counties \citep{wooldridge2010econometric, angrist2009mostly}.

\begin{equation}
\label{eq:logit}
\begin{aligned}
\Pr(Y_{itm} =1 )
= \text{logit}^{-1}\!\biggl(
\beta_{0}
&+ \beta_{1}\,\texttt{UnemploymentRate}_{itm}
+ \beta_{2}\,\frac{\texttt{LaborForce}_{itm}}{1000}\\
&+ \beta_{3}\,\texttt{UnemploymentRate}_{itm}\times\frac{\texttt{LaborForce}_{itm}}{1000} \\
&+ \beta_{4}\,\frac{\texttt{ACS5}_{itm}}{1000}
+ \beta_{5}\,\texttt{Rural}_{i}
+ \beta_{6}\,\texttt{TweetVol}_{itm} \\
&+ \gamma_{t}
+ \delta_{m}
+ \theta_{s(i)}
\biggr),
\end{aligned}
\end{equation}
where the index $s(i)$ denotes the state of county~$i$, $\gamma_{t}$ are the year fixed effects, $\delta_{m}$ the month fixed effects, and $\theta_{s(i)}$ are the state fixed effects that absorb time-invariant differences across states.
To assess whether our results are sensitive to the dichotomization of the sentiment measure into a binary outcome, we also estimate a traditional OLS model (see equation~\eqref{eq:ols}) following the same nested specification strategy described above.
 The results are presented in the Supplementary Material in Table~\ref{tab:jobsat_OLS} and they confirm the results of the logit model  in Table~\ref{tab:jobsat}.
All inference is based on heteroskedasticity-robust standard errors clustered at
the county level, which accounts for arbitrary within-county correlation in the
error structure over time. This correction is standard in panel-data settings
with many groups and relatively long time series, and ensures valid
asymptotic inference under general forms of within-cluster dependence
\citep{white1980heteroskedasticity, liang1986longitudinal, cameron2015practitioner}.

\subsection{Analysis of the results} 
The logistic regression estimates reported in Table~\ref{tab:jobsat} indicate a highly
stable and theoretically coherent set of relationships between local
socioeconomic conditions and expressed job-satisfaction sentiment. Across all
specifications, the unemployment rate emerges as the strongest and most robust
predictor: higher unemployment is consistently associated with a lower
probability that county-month sentiment is positive, even after conditioning on
income, rurality, tweet volume, labor-force size, and a full set of temporal and
state fixed effects. The sequential inclusion of ACS5 income, the interaction between unemployment and labor force, and rurality attenuates the unemployment coefficient only modestly. All three variables enter with statistically significant signs: higher income is associated with higher job satisfaction, aligning with expectations; more rural counties exhibit significantly lower sentiment net of labor-market conditions and socioeconomic controls; and the negative interaction term indicates that the adverse association between unemployment and job satisfaction is stronger in counties with larger labor forces. Notice that the impact of rurality on job satisfaction is negative, while the effect of rurality on life satisfaction is, in most of the literature, positive \citep{iacus2025b}.

The average marginal effects reported in Table~\ref{tab:ame_jobsat} provide a
direct quantification of the substantive importance of each predictor in the
logit model. Higher unemployment rates are associated with a lower probability
that job-satisfaction sentiment is positive, even after conditioning on
socioeconomic characteristics and fixed effects, although the magnitude of the
effect is moderate when expressed in probability units. County income
(ACS5) exhibits a positive and statistically precise relationship with
job-satisfaction sentiment, indicating that counties with higher economic
resources tend to display higher levels of expressed job satisfaction.
Rurality remains one of the strongest predictors: moving one step toward a more
rural classification reduces the probability of positive sentiment by nearly one
percentage point on average, net of controls. Labor-force size has a small but
positive marginal effect, consistent with slightly higher job satisfaction in
larger local labor markets, while the interaction between unemployment and labor
force contributes an additional, though modest, strengthening of the negative
impact of unemployment in these larger markets. Tweet volume enters with a small but
negative marginal effect, suggesting that months in which more users discuss
work topics tend to coincide with more negative sentiment overall. On the whole, the AMEs corroborate the core conclusions from the full
logit specification, while translating them into effect sizes that are directly
interpretable on the probability scale.

\begin{figure}[!h]
 \caption{Average difference $\Delta P = P(\texttt{jobsat}>0\, |\, \texttt{rural} = r) - P(\texttt{jobsat}>0\, | \,\texttt{rural} = 1)$, between each rurality level ($r=2,\ldots,9$) and the most urban counties ($r=1$), plotted across unemployment rates. Predicted probabilities are obtained from the logit model in equation~\eqref{eq:logit} and averaged over the empirical distribution of year, month, and state fixed effects.}
    \label{fig:jobsat_free}
    \centering
    \includegraphics[width=0.85\linewidth]{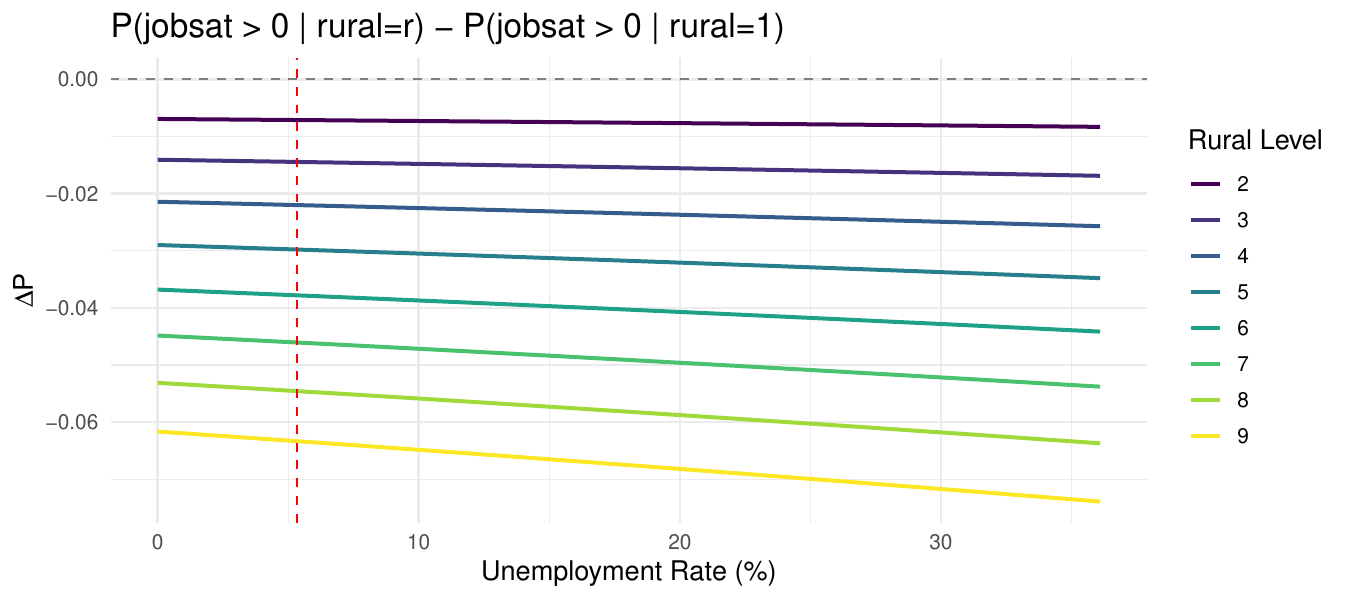}
\end{figure}

Figure~\ref{fig:jobsat_free} visualizes the relationship between unemployment and job satisfaction across the rural--urban continuum. Predicted probabilities are obtained from equation~\eqref{eq:logit} by evaluating the model over a grid of unemployment rates (0--30\%) and rurality levels (1--9). Continuous covariates (labor force, ACS5 income, and tweet volume) are held at their median values. To account for temporal and spatial heterogeneity, predictions are averaged over the empirical distribution of month and state fixed effects. The final curves represent  
the average difference 
$\Delta P = P(\texttt{jobsat}>0 \mid \texttt{rural}=r) - P(\texttt{jobsat}>0 \mid \texttt{rural}=1)$ 
between each rurality level ($r=2,\ldots,9$) and the most urban counties ($r=1$).

Figure~\ref{fig:jobsat_free} shows that the curves converge at low unemployment rates, indicating a partial rural catch-up, but diverge  as unemployment rises, revealing an increasing rural disadvantage under weaker labor-market conditions.

\begin{table}[!htbp] 
\centering 
\caption{Sequential logit specifications for the probability that county-month 
job-satisfaction sentiment is positive ($P(\texttt{jobsat} > 0)$). 
Model~(1) includes unemployment, labor force, tweet volume, and year and month fixed effects; 
Model~(2) adds ACS5 income; Model~(3) adds an unemployment time labor force interaction;  Model~(4) further includes rurality and state fixed effects. Fixed effects are included but omitted from the table.}
  \label{tab:jobsat} 
  \scriptsize
\begin{tabular}{lcccc}
   \tabularnewline \midrule \midrule
   Dependent Variable: & \multicolumn{4}{c}{$P(\texttt{jobsat}>0)$}\\
   Model:                                                      & (1)                     & (2)                     & (3)                     & (4)\\  
   \midrule
   \emph{Variables}\\
   Unemployment rate                                           & -0.0764$^{***}$         & -0.0423$^{***}$         & -0.0299$^{***}$         & -0.0101$^{*}$\\   
                                                               & (0.0047)                & (0.0050)                & (0.0051)                & (0.0052)\\   
   Labor force (thousands persons)                             & 0.0027$^{***}$          & 0.0016$^{***}$          & 0.0039$^{***}$          & 0.0033$^{***}$\\   
                                                               & (0.0007)                & (0.0006)                & (0.0007)                & (0.0007)\\   
   Valid jobsat tweets                                         & -0.0002$^{***}$         & -0.0001$^{***}$         & -0.0001$^{***}$         & -0.0001$^{***}$\\   
                                                               & ($3.14\times 10^{-5}$)  & ($3.08\times 10^{-5}$)  & ($3.17\times 10^{-5}$)  & ($3.3\times 10^{-5}$)\\    
   ACS5 income (thousands USD)                                 &                         & 0.0221$^{***}$          & 0.0210$^{***}$          & 0.0140$^{***}$\\   
                                                               &                         & (0.0018)                & (0.0016)                & (0.0016)\\   
   Unemployment rate $\times$ Labor force (thousands persons)  &                         &                         & -0.0003$^{***}$         & -0.0003$^{***}$\\   
                                                               &                         &                         & ($4.84\times 10^{-5}$)  & ($4.38\times 10^{-5}$)\\    
   Rurality                                                    &                         &                         &                         & -0.0871$^{***}$\\   
                                                               &                         &                         &                         & (0.0063)\\   
   \midrule
   \emph{Fixed-effects}\\
   year                                                        & Yes                     & Yes                     & Yes                     & Yes\\  
   month                                                       & Yes                     & Yes                     & Yes                     & Yes\\  
   State                                                       &                         &                         &                         & Yes\\  
   \midrule
   \emph{Fit statistics}\\
   Observations                                                & 282,079                 & 282,065                 & 282,065                 & 281,941\\  
   Pseudo R$^2$                                                & 0.39363                 & 0.39950                 & 0.40079                 & 0.42067\\  
   Squared Correlation                                         & 0.44230                 & 0.44420                 & 0.44410                 & 0.46273\\  
   AIC                                                         & 181,018.9               & 179,263.8               & 178,881.2               & 173,009.2\\  
   BIC                                                         & 181,282.6               & 179,538.1               & 179,166.1               & 173,810.9\\  
   \midrule \midrule
   \multicolumn{5}{l}{\emph{Clustered standard-errors  (by county) in parentheses}}\\
   \multicolumn{5}{l}{\emph{Signif. Codes: ***: 0.01, **: 0.05, *: 0.1}}\\
\end{tabular}
\end{table}

\begin{table}[!ht]
\centering 
\caption{Average marginal effects from the logit model in equation~\eqref{eq:logit}. 
Effects are averaged across county-month observations; year, month, and state fixed effects are 
included but not shown. Standard errors clustered at the county level.}
  \label{tab:ame_jobsat} 
  \small
\begin{tabular}{@{\extracolsep{5pt}} lcccc} 
\\[-1.8ex]\hline 
\hline \\[-1.8ex] 
Variable & AME & SE & z & $p$-value \\ 
\hline \\[-1.8ex] 
Unemployment rate & -0.00208 & 0.00046 & -4.49 & 6.966e$^{-06}$ \\ 
Labor force (thousands persons) & 0.00016 & 0.00004 & 3.78 & 0.0001567 \\ 
ACS5 income (thousands USD) & 0.00131 & 0.00015 & 8.84 & 9.217e$^{-19}$ \\ 
Rurality & -0.00809 & 0.00058 & -13.87 & 9.434e$^{-44}$ \\ 
Valid jobsat tweets & -0.00001 & 0.00000 & -3.24 & 0.001214 \\ 
Unemp. $\times$ labor force (interaction) & -0.00002 & 0.00000 & -5.86 & 4.654e$^{-09}$ \\ 
\hline \\[-1.8ex] 
\end{tabular} 
\end{table}

\clearpage

\section{Discussion: what rural--urban convergence, if any?}

Between 2013 and 2023, U.S. unemployment rates experienced a steady decline, interrupted only during the pandemic years (\cite{BLS2025}). Over the same period, however, rural--urban inequality increased in multiple domains. Income gaps widened (\cite{hardy2025}), employment opportunities grew more unevenly across space (\cite{dumont2023}), and poverty rates in nonmetropolitan areas remained persistently higher than in metropolitan ones (\cite{usda2024}). In short, the economic expansion of the last decade did not generate broad-based rural--urban convergence in objective economic conditions.
This period also coincided with pronounced job polarization, characterized by an expansion of employment at both the low-skill and high-skill ends of the labor market and a relative decline in mid-skill occupations. These developments reflect deeper structural transformations associated with technological change, task reallocation, and the declining demand for routine work (\cite{acemoglu2011, acemoglu2019, autor2019}). Importantly, these dynamics have not affected rural and urban areas symmetrically. Earnings dispersion, particularly by education level, increased more in urban labor markets than in rural ones (\cite{autor2019}). As \cite{hardy2025} observe, the rural--urban divergence in household income is driven largely by widening gaps at the top of the distribution, with only limited convergence at the bottom.
These differences help contextualize the patterns we observe in our job-satisfaction indicator.
 Declining unemployment rates are associated with higher job-satisfaction sentiment across all counties, and the overall urban-rural ranking remains unchanged: more urban counties exhibit higher average sentiment than more rural ones. However, the magnitude of this rural disadvantage is not constant. Figure~\ref{fig:jobsat_free} shows that the predicted gap in job satisfaction narrows substantially at low unemployment rates, whereas it widens when unemployment is higher. This form of conditional convergence is consistent with a setting in which labor-market tightness disproportionately benefits residents of rural counties, even as average levels of job satisfaction remain higher in urban areas.

One possible explanation relates to differences in the evolution of earnings dispersion. If relative income concerns \citep{duesenberry1949} play a role in shaping job satisfaction, rising inequality within urban counties may partly offset the positive effect of declining unemployment on perceived job quality. Rural counties, where earnings dispersion has increased less, may experience larger subjective gains when labor markets tighten. While our empirical design does not directly test this mechanism, the conditional convergence pattern we estimate is directionally consistent with these documented structural differences.
This stands in contrast to broader life-satisfaction and happiness indicators, which in our related HFGI analyses exhibit weaker or inconsistent associations with relative income dynamics \citep{deaton2013, iacus2025b}. Job satisfaction, being more domain-specific and closely tied to labor-market conditions, may be more sensitive to relative economic position.
Finally, the conditional improvement in predicted job satisfaction among rural counties aligns with aspects of \cite{autor2019}'s interpretation of changing opportunities for less-educated workers. The slowdown in migration of low-skill labor into urban areas may place upward pressure on urban wages, but demographic change, especially population aging, has also increased demand for labor-intensive, lower-skill services in nonmetropolitan areas.
These shifts may be expanding employment opportunities in rural counties in ways not fully captured by standard income measures. The patterns in our data, i.e., higher predicted gains in job satisfaction  for rural counties under tight labor markets, are consistent with this interpretation and suggest that perceived job conditions may be improving more quickly in lower-density labor markets when unemployment is low.

\section{Conclusions}
The econometric analysis in this work shows that the rural--urban gap in job satisfaction is not fixed, but varies systematically with local labor-market conditions. Higher unemployment lowers job satisfaction across all counties, and this negative effect is stronger in rural areas. When unemployment is low, however, the estimated rural disadvantage \textit{narrows}. This pattern reflects a form of conditional convergence in \textit{perceived} job quality.
In this respect, these findings complement Autor's evidence by showing that, even as rural-urban income disparities widen, perceived job satisfaction converges conditionally under tight labor markets. Subjective assessments of job quality improve relatively more in rural counties when unemployment is low, revealing a dimension of well-being in which residents of more rural counties appear to catch up.
 This conditional and perception-based convergence represents a new and economically meaningful feature of the U.S. labor market.
Although income gaps continue to widen, our econometric results suggest that rural labor markets realize larger gains in subjective job satisfaction when labor demand is strong. Policies that help sustain tight labor-market conditions in non-metropolitan areas, such as targeted job creation, support for care and service sectors, and investments that stabilize local employment, could therefore yield meaningful improvements in perceived job quality. Because job satisfaction is linked to retention, productivity, and migration decisions, such policies may strengthen rural labor markets even without immediate income convergence.

\bibliography{sample}

\clearpage

\section*{Supplementary Material}
\renewcommand{\thefigure}{S\arabic{figure}}
\renewcommand{\thetable}{S\arabic{table}}
\renewcommand{\theequation}{S\arabic{equation}}
\setcounter{figure}{0}
\setcounter{table}{0}
\setcounter{equation}{0}

\subsection*{Large language model fine-tuning and construction of social media based indicators}

We summarize here how the tweet-level indicator of \emph{job satisfaction} was derived using fine-tuned large language models, and further aggregated to the county level. 
A full methodological account of the broader {Human Flourishing Geographic Index (HFGI)} dataset, including additional well-being dimensions, validation benchmarks, and reproducibility resources is provided in \cite{iacus2025}. 
Here we describe only the indicator used in this work:  \texttt{jobsat}.

\subsection*{Fine-tuning and classification pipeline}

We trained and validated a fine-tuned large language model (LLM) to classify short social-media texts according to multiple flourishing-related constructs. 
The training corpus comprised 4,581 manually annotated tweets, selected from an initial pool of 10,000 messages sampled across topics, years, and geographies to ensure linguistic diversity. 
Each tweet was independently labeled by human coders following standardized dimension-specific guidelines. 
For  \texttt{jobsat}, coders indicated whether the tweet referred to the construct and, if so, whether the expressed tone was \textit{low}, \textit{medium}, or \textit{high}. 
These labels were used to fine-tune an instruction-following generative model (LLaMA family), optimizing cross-entropy loss across the three classes.

The fine-tuned model produced a JSON-formatted classification for each tweet, returning both the target label and confidence score. 
Evaluation on a held-out test set yielded accuracies around 0.87 depending on the dimension, with consistent precision-recall balance. 
Post-hoc validation involved manual review of ambiguous cases to verify conceptual alignment with the human flourishing framework (see \cite{iacus2025} for details).

\subsection*{Scaling and county-level aggregation}
We later transformed the three-level scale (``low", ``medium", ``high'') as: $-1$ (\textit{negative}), $+0.5$ (\textit{somewhat positive}), $+1$ (\textit{positive}) and set $0$ if a tweet does not contain expressions related to job satisfaction (\textit{not-present}). Notice that low maps to the opposite concept. For example ``job satisfaction" = ``low" means ``unsatisfied on the job".

Each classified tweet was geolocated to a U.S.\ county using the Harvard CGA pipeline.  Tweets with ambiguous or low-confidence geotags were excluded. 
For every county-month-dimension triple, we computed the mean and standard deviation per geographical region (county/state) and temporal frequency (month/year).

\subsection*{Labor-market variable selection}
For county-level labor-market indicators we rely on those that can be obtained from the U.S.\ Bureau of Labor Statistics (BLS) \citep{BLSLAUS} Local Area Unemployment Statistics (LAUS) program. Table~\ref{tab:jobsat_correlations} shows that our job satisfaction \texttt{jobsat} indicator is  correlated more to the \texttt{Labor force} indicator and to the \texttt{Unemployment rate} at county level than to the other labor market indicators, so we selected these two measures as control variables. Table~\ref{tab:jobsat_correlations} shows that our job-satisfaction indicator \texttt{jobsat} is moderately negatively correlated with the \texttt{Unemployment rate} and weakly positively correlated with the \texttt{Labor force} measure at the county level, while its associations with other covariates are smaller in magnitude. We therefore selected the \texttt{Unemployment rate} and the \texttt{Labor force} as our primary labor-market controls in the econometric analysis. As the variable \texttt{Labor force} is almost perfectly correlated with the \texttt{Population} indicator, we include only the former in the models. The rest of the variables are defined in Section~\ref{sec:data}.

\begin{table}[ht]
\caption{Pairwise correlations between \texttt{jobsat} and other socio-economic covariates.}
\label{tab:jobsat_correlations}
\centering
\scriptsize
\begin{tabular}{llllllll}
 \texttt{Employed} & \texttt{Labor force} & \texttt{Unemployed} & \texttt{Unemployment rate} & \texttt{Rural} & \texttt{Population} & \texttt{TweetVol} & \texttt{ACS5} \\ 
\hline 0.04 & 0.04 & -0.02 & -0.31 & -0.13 & 0.03 & -0.01 & 0.13 \\ 
\end{tabular}
\end{table}

\subsection*{Alternative OLS specification}
To assess whether our results from the model in equation~\eqref{eq:logit} in Section~\ref{sec:model} are sensitive to the dichotomization of the sentiment measure into a binary outcome, we also estimate a traditional OLS model as in equation~\eqref{eq:ols} below, following the same nested specification strategy described in Section~\ref{sec:model}.
\begin{equation}
\label{eq:ols}
\begin{aligned}
\text{jobsat}_{itm}
=
\beta_{0}
&+ \beta_{1}\,\texttt{UnemploymentRate}_{itm}
+ \beta_{2}\,\frac{\texttt{LaborForce}_{itm}}{1000}\\
&+ \beta_{3}\,\texttt{UnemploymentRate}_{itm}\times\frac{\texttt{LaborForce}_{itm}}{1000} \\
&+ \beta_{4}\,\frac{\texttt{ACS5}_{itm}}{1000}
+ \beta_{5}\,\texttt{Rural}_{i}
+ \beta_{6}\,\texttt{TweetVol}_{itm} \\
&+ \gamma_{t}
+ \delta_{m}
+ \theta_{s(i)}
+ \varepsilon_{itm},
\end{aligned}
\end{equation}
where $\varepsilon_{itm}$ is an idiosyncratic error term with zero mean, representing unobserved determinants of $\text{jobsat}_{itm}$ conditional on the covariates and fixed effects, i.e.\ $\mathbb{E}[\varepsilon_{itm}\mid X_{itm},\gamma_{t},\delta_{m},\theta_{s(i)}]=0$.

\begin{table}[!ht] 
\centering 
\caption{OLS regression results for county-month job-satisfaction sentiment
(\texttt{jobsat}). The four models follow the same nested structure as in
Table~\ref{tab:jobsat}. All specifications include year and month fixed
effects; Model~(4) additionally includes state fixed effects. Standard errors
are heteroskedasticity-robust and clustered at the county level.}
  \label{tab:jobsat_OLS} 
  \scriptsize
\begin{tabular}{@{\extracolsep{5pt}}lcccc} 
\\[-1.8ex]\hline 
\hline \\[-1.8ex] 
 & \multicolumn{4}{c}{\textit{Dependent variable:}} \\ 
\cline{2-5} 
\\[-1.8ex] & \multicolumn{4}{c}{\texttt{jobsat}} \\ 
\\[-1.8ex] & (1) & (2) & (3) & (4)\\ 
\hline \\[-1.8ex] 
 Unemployment rate & $-$0.01300$^{***}$ & $-$0.01144$^{***}$ & $-$0.01142$^{***}$ & $-$0.00727$^{***}$ \\ 
  & (0.00055) & (0.00056) & (0.00057) & (0.00049) \\ 
  Labor force (thousands persons) & 0.00004$^{***}$ & 0.00002 & 0.00002 & 0.00004$^{**}$ \\ 
  & (0.00002) & (0.00001) & (0.00002) & (0.00001) \\ 
  ACS5 income (thousands USD) &  & 0.00081$^{***}$ & 0.00081$^{***}$ & 0.00019$^{*}$ \\ 
  &  & (0.00011) & (0.00011) & (0.00011) \\ 
  Rurality &  &  &  & $-$0.00749$^{***}$ \\ 
  &  &  &  & (0.00055) \\ 
  Valid jobsat tweets & $-$0.00001$^{***}$ & $-$0.00001$^{***}$ & $-$0.00001$^{***}$ & $-$0.00001$^{***}$ \\ 
  & (0.000002) & (0.000002) & (0.000002) & (0.000002) \\ 
  Unemployment rate $\times$ Labor force (thousands) &  &  & $-$0.0000004 & $-$0.000001 \\ 
  &  &  & (0.000001) & (0.000001) \\ 
  Constant & $-$0.06715$^{***}$ & $-$0.11624$^{***}$ & $-$0.11632$^{***}$ & $-$0.07870$^{***}$ \\ 
  & (0.00510) & (0.00816) & (0.00820) & (0.01031) \\ 
 \hline \\[-1.8ex] 
Akaike Inf. Crit. & 3010.23 & 2466.65 & 2468.48 & -8807.15 \\ 
Year \& month FE & Yes & Yes & Yes & Yes \\ 
State FE & No & No & No & Yes \\ 
Clustered SE (county) & Yes & Yes & Yes & Yes \\ 
Observations & 282,079 & 282,065 & 282,065 & 282,065 \\ 
R$^{2}$ & 0.62892 & 0.62963 & 0.62963 & 0.64427 \\ 
Adjusted R$^{2}$ & 0.62889 & 0.62960 & 0.62960 & 0.64417 \\ 
\hline 
\hline \\[-1.8ex] 
\textit{Note:}  & \multicolumn{4}{l}{$^{*}$p$<$0.1; $^{**}$p$<$0.05; $^{***}$p$<$0.01} \\ 
 & \multicolumn{4}{l}{Heteroskedasticity-robust standard errors clustered at the county level (StateCounty).} \\ 
\end{tabular} 
\end{table} 

\end{document}